\magnification = \magstep1
\parindent = 0pt
\centerline{\bf A first sample of faint radio sources with virtually
complete
redshifts}
\centerline{\bf I. Infrared images, the Hubble diagram, and the alignment
effect}
\vskip 0.4truein
\centerline{\bf Stephen Eales$^1$, Steve Rawlings$^2$, Duncan Law-Green$^4$,
Garret Cotter$^3$}
\centerline{\bf and Mark Lacy$^2$}
\vskip 0.4truein
1 Department of Physics and Astronomy, University of Wales Cardiff,
P.O. Box 913, Cardiff CF2 3YB
\bigskip
2 Department of Astrophysics, University of Oxford, Nuclear and 
Astrophysics Laboratory, Keble Road, Oxford, OX1 3RH
\bigskip

3 Mullard Radio Astronomy Observatory, Cavendish Laboratory, Madingley
Road, Cambridge CB3 0HE

\bigskip

4 Nuffield Radio Astronomy Laboratories, Jodrell Bank, Macclesfield,
Cheshire SK11 9D11
\bigskip

\centerline{\bf Abstract}
\bigskip
We have obtained redshifts and infrared images for a sample of faint
B2/6C radio sources whose fluxes are about six times fainter
than those of sources in the bright 3C sample. We now have unambiguous
redshifts
for 90\% of the sources, making this
the first faint radio sample with such complete redshift information.
We find that the infrared Hubble diagrams (K versus z) of the 
3C sample and the B2/6C sample are similar at a low redshift ($\rm z < 0.6$),
but by a redshift of $\rm z \sim 1$ the 6C/B2 galaxies are 
on average $\simeq$0.6 mags
fainter in the K-band than 3C galaxies at the same redshift.
This suggests that the bright K-magnitudes of 3C galaxies at $\rm z \sim 1$
are not the result of stellar evolution, but of a correlation between infrared
and radio luminosity. We also find that the infrared stuctures of
B2/6C galaxies at $\rm z \sim 1$ are less clearly
aligned with their radio structures than 3C galaxies at
this redshift, implying that the strength of
the alignment effect depends on radio luminosity.
Finally, above a redshift of 2 we find that 
the dispersion in the K-z relation of the B2/6C sample
is
$\rm \simeq 2.7$ times greater
than that at low redshift, a result which is expected
if
at these redshifts we are beginning to penetrate into
the epoch in which radio galaxies formed.

\vfill
\eject

\centerline{\bf 1. Introduction}

\bigskip
\parindent = 20pt

Radio galaxies appear very different at low and high redshifts.
In the local universe radio galaxies are always giant elliptical
galaxies with a small spread of absolute magnitude (Laing, Riley
\& Longair 1983). In the high-redshift
universe the optical image of a typical radio galaxy consists of a series
of knots strung out along the radio axis, nothing like the image
expected for a giant elliptical, with the alignment suggesting a connection
between the optical and radio emission (Chambers, Miley \& van Breugel 1987;
McCarthy et al. 1987; Longair, Best \& Rottgering 1995).  
Other differences are that high-redshift radio galaxies (henceforth
HZRG's) have brighter
near-infrared magnitudes and bluer optical-infrared colours than are
expected from the properties of their low-redshift counterparts
(Lilly \& Longair 1984), higher emission-line luminosities
(Rawlings et al. 1989; Jackson \& Rawlings 1996; Eales
\& Rawlings 1993, 1996), and smaller radio sizes (Eales 1985c); 
radio galaxies at high redshift are often surrounded by
extensive ($D \sim 100$ kpc) line-emitting nebulae (McCarthy,
van Breugel \& Kapahi 1991) that are not seen around those at
low redshift; and finally the optical emission from 
HZRG's is often polarised, with the electric vector perpendicular
to the radio axis, again suggesting a connection between the radio and optical
emission (Tadhunter et al. 1992).

It is tempting to attribute these differences to cosmic evolution.
For example, a natural explanation of the brighter near-infrared
magnitudes of HZRG's
is the expected cosmic evolution of the stellar population
of a radio galaxy.
A present-day giant elliptical galaxy contains lots of stars but very little
current star formation, suggesting that a large burst of star formation
must have happened at some point in its past. If so, 
the galaxy should gradually have declined in brightness after the
burst because
of the evolution of the main-sequence turnoff mass.
Lilly and Longair (1984) showed that the amount of evolution present in
the radio galaxy K-z diagram is in agreement with detailed models 
of this effect. This was a very satisfying result because, as low-redshift
radio galaxies do contain stars, it would have been surprising if the
effects of stellar evolution had not been seen for radio galaxies at
$\rm z \sim 1$, which for a universe with $\rm \Omega_0 = 1$ is
an epoch 65\% of the way back to the beginning of the universe.
The other differences can also be explained, although
usually in a more ad hoc and often unsatisfactory way. The optical
structures of HZRG's, for example, do look
remarkably like simulations arising from models of galaxy formation based
on hierarchical clustering (e.g. Fig. 3 of Baron \& White 1987),
but this does not explain why the optical structures should 
be aligned
with the radio axis [Eales (1992) gives a possible explanation]. 

Notwithstanding this temptation,
one
selection effect makes it impossible to be sure that any difference
is caused by cosmic evolution. Almost all the results above were
obtained from observations of the bright 3C sample of radio sources 
(Laing, Riley and Longair 1983). Since 3C is
a flux-limited sample, radio luminosity and redshift are tightly correlated 
(Fig. 1), which means that
that any difference in the properties of low- and high-redshift radio
galaxies could equally well be due to the difference in their 
radio luminosities as to
the difference in their redshifts. There are also often explanations
of the observed differences based on luminosity that are equally as
plausible as those based on cosmic evolution. For example, the difference
in emission-line luminosities seems to be most naturally explained by  
the idea that the emission-line gas is being photoionized by continuum
emission from the active nucleus; both the emission-line
luminosity and the radio luminosity then scale with the energy output
of the active nucleus (Rawlings \& Saunders 1991), with the high emission-line
luminosities of the HZRG's being caused
by
their high radio luminosities rather than 
by their redshifts. 

This is especially true in the case of the alignment effect, because
although it is possible to think of explanations based on cosmic evolution
or even simply on the effect of the redshift (if the alignment effect is
primarily a $UV$ phenomenom, it will be more obvious to optical observations
as the redshift increases), 
most of the suggested explanations are based on the high
radio luminosities of the high-redshift 3C galaxies not on their
redshifts. For example, the intensity and direction (E-vector perpendicular
to the radio axis) of the polarized optical emission from a typical
HZRG
suggests a model in which 
the HZRG
contains a quasar that we cannot see directly but which is
emitting optical radiation along the radio axis in the plane of the sky,
with the aligned optical component being light from the quasar
that is scattered towards us  
(Tadhunter, Fosbury
\& di Serego Alighieri 1988). In this model the brightness of
this extended emission, and thus the quality of the alignment
between the radio and optical axes, will depend on the brightness
of the quasar, and thus on the energy output of the central engine.
Since the radio luminosity will also depend on this, this hypothesis
makes the prediction that the strength of the alignment effect depends
on the radio luminosities of the galaxies not on their redshifts.
A similar prediction comes from the hypothesis that the alignments are
caused by star formation being triggered along the axis of a radio
source by the bowshock of the expanding source (McCarthy et al.
1987).
Dunlop \& Peacock (1993) have found evidence that
the alignment effect is indeed related to radio luminosity rather than
to redshift. They find that the K-band structures of 3C
radio galaxies at $\rm z \simeq 1$ are closely aligned with the
radio axes, but that the K-band structures of sources from a faint
radio sample whose K-band magnitudes suggest they are at a similar
redshift to the 3C galaxies are not aligned with the radio axes.

As the last result illustrates, the way to 
disentangle the effects of redshift and radio luminosity
is to compare the properties of a faint radio sample with those of a bright
radio sample such as 3C.
Although radio luminosity and redshift will
be tightly correlated within the faint sample, when the
two samples are considered together there will be a much larger
range of radio luminosity at constant redshift 
than is available within a single flux-limited sample
(the range of luminosity
is given roughly by the ratio of the flux limits of the two
samples),
and similarly there will be a larger range of redshift at constant
radio luminosity. 
Although redshifts have been measured for large numbers of HZRG's
in faint radio samples (there are over 100 radio galaxies known at
$\rm z > 2$, for example---McCarthy 1993), most have been found using
radio criteria designed to preferentially select high-redshift sources,
and thus they form a biased sample. To give one example, McCarthy et al. (1991)
observed a large number of sources selected from the Molongo Radio Catalogue
with spectral indices greater than 0.9. The advantage of using this
spectral-index criterion is that it is effective (for reasons that are
not understood) at selecting the most luminous sources, and thus the
ones that are at the highest redshifts. These also are the ones for which
it is easiest to measure redshifts (line luminosity is correlated with
radio luminosity [\S 5.1], and the high redshifts put bright lines in accessible
parts of the spectrum). The disadvantage is that introducing this criterion
biases the sample in ways that are not properly undestood.
The practical difficulty in the way of measuring redshifts for {\it all}
the sources in faint radio samples 
has been the exhorbitant amount of telescope time needed, 
a problem that appears particularly daunting
when one considers it took
30 years of observational work to do the same for 3C. 

Over the last decade we and collaborators
have been gradually obtaining optical, infrared
and radio images, and, most importantly, redshifts for the sources
in the faint samples of Allington-Smith (1982) and Eales (1985a), 
samples drawn respectively from
the 408-MHz B2 survey
and from the 151-MHz 6C survey,
and both with
flux limits about six times fainter
than that of 
3C. 
Radio samples selected at different frequencies contain different
proportions of sources from the different radio morphological types (e.g.
Peacock
and Wall 1981),
but since 3C was also 
carried out at a low frequency (178 MHz), comparisons of the faint samples
with 3C should not suffer from this problem. 
Table 1 shows the parameters of the original samples.
The two faint samples contain 
many sources in common, since they were selected in almost the same area
of sky and have similar flux limits. 
The total number of sources in
the original samples was 84. Three of the sources are now known to the
result of the confusion of two or more sources within the telescope's
beam, with the fluxes of the separate sources not falling within the
flux limits of the samples, and the flux originally given for one
source is now known to be wrong, which leaves 80 sources remaining within the
samples.

We have now completed our observational
study of these samples and have redshifts and infrared magnitudes for
a large fraction of the sources, as well as many new radio maps and a
limited amount of optical imaging data. 
The crucial data are the redshifts. Here we have redshifts for 72
(90\%) of the 80 sources. Of the sources without redshifts, one is
too close to a star for us to obtain either an image or a redshift,
two sources are fairly bright galaxies ($\rm R \sim 20$), for which
we believe we can reliably estimate redshifts (see below), one
is a BL Lac object, and four
appear too faint for us to be able to measure redshifts with 4m-class
telescopes, despite several attempts.
As far as
we are aware, these samples are the first faint samples with
such complete redshift information.
Apart from three low-redshift galaxies, two quasars, and
the source close to the star, we have
infrared photometry, measured either with a single-element detector
or with a camera,
for all of the sources in the samples.
The way we intend to present the data is as follows. 
The new radio maps have already
been published (Naundorf et al. 1991; Law-Green et
al. 1995a). This paper will present the infrared images and
a subsequent paper (Rawlings et al. 1996---Paper II) will contain
the optical spectra. 
A final paper (Paper III) will deal with some of the technical
issues concerning the samples (some of the radio flux scales
have changed since the original papers, for example) and will
provide a compilation of all the data and a reference list.

At the same time, and often in the same papers, we will compare
the data for the faint samples with that for 3C, in an attempt
to determine whether the differences between the low-redshift
and high-redshift 3C sources are genuinely caused by cosmic evolution.
We have already used the faint samples to show that the physical
sizes of double radio sources are smaller at high redshift as the
result of cosmic evolution rather than being the result of a luminosity
effect (Neeser et al. 1995). In this paper we turn our attention to
Lilly and Longair's (1984) discovery that high-redshift 3C galaxies are
systematically brighter in the K-band than expected from
the absolute magnitudes of radio galaxies at low redshift. Although
this result is nicely in accord with one's preconceptions about galaxy
evolution---stellar evolution should make galaxies brighter in the past!---in
this paper we show that it is the high radio luminosities of the high-redshift
3C galaxies that are responsible for their bright K-band magnitudes.
We will also consider the question of whether the strength of
the alignment effect depends on radio luminosity.
Paper II will address the question of whether the emission-line
luminosity of a radio galaxy depends on its redshift or its radio luminosity
and also the fundamental question of whether the comoving space density
of radio galaxies reaches a maximum at some redshift.

The layout of this paper is as follows. We present the data in \S 2.
We compare the Hubble diagrams of the faint samples and of the bright
3C sample in \S 3, and compare the alignment effect
in the different samples in \S 4. We discuss the consequences of these
comparisons in \S 5.
We assume a Hubble constant of 
50 km s$^{-1}$ Mpc$^{-1}$.

\bigskip
\centerline{\bf 2. Observations and Data Reduction}
\bigskip
\centerline{\bf 2.1. The Observations}
\bigskip
Most of the observations were made during three observing runs. In
1994 Jan 16-18, we used the IRCAM infrared
camera on the United Kingdom Infrared Telescope (UKIRT)
to observe some of the sources in the infrared K-band (2.2 $\mu$m). 
IRCAM consists
of a 58 $\times$ 62 element
SBRC infrared array and has a plate scale of 0.62 arcsec pixel$^{-1}$.
In 1995 Jan 18-20 we used the REDEYE camera on the Canada-France-Hawaii
Telescope (CFHT) to observe sources in the K$'$-band.
REDEYE
consists of four 128 $\times$ 128 element HgCdTe mosaics butted
together (there is a small gap between the two lower
and two upper arrays) with a plate
scale of 0.5 arcsec pixel$^{-1}$. The
K$'$-band is similar to the K-band but is shifted
to a slightly shorter
wavelength in order to reduce the noise caused by the
rapidly increasing thermal background in the long-wavelength part
of the K-band filter. On
1995 Jan 25 we used the new IRCAM3 camera on the UKIRT to
make observations in the K-band. IRCAM3 consists 
of a 256 $\times$ 256 element SBRC infrared array and has a plate scale of 
0.286 arcsec pixel$^{-1}$. 
A handful of galaxies were observed with
IRCAM or IRCAM3 during other regular observing runs or during
service runs.
We observed many of the galaxies
in our sample with two or more of these cameras, which gives us a useful
check on the photometric accuracy of our observations (\S 2.4).
Table 2 lists the sources we observed and the instruments we used.

In all cases we used the standard technique of observing each galaxy
at nine positions on a 3 $\times 3$ grid with a spacing between gridpoints of 8
arcsec. As long as the background emission is either
mainly produced by the sky, or is from
the telescope but is spatially flat (Bunker 1996), this is a useful procedure,
because
an accurate flat field can be produced from the data themselves (\S 2.2).
We typically integrated for about three minutes per position, and in many
cases we repeated the sequence of nine observations several times. 
Total exposure
times are given in Table 2. We photometrically calibrated our data
by observations of standards from the list of Elias (1982), mainly
7th-magnitude stars, and from the UKIRT list of faint 
standards (10-14th magnitude), which appears on the UKIRT pages
of the worldwide web. Note that the REDEYE observations were made
through a K$'$-filter but were calibrated using K-band standards. This
means that to obtain proper K-band magnitudes from these
observations we had to make a colour-correction. This is described
in \S 2.2.

Since the size of IRCAM was not large enough in many cases for us
to
obtain astrometric information from objects visible on the final
image, for most of the
galaxies we used the following astrometric procedure.
For each galaxy we first obtained an image
of a nearby star with known position, then
moved
the telescope to the position of the radio galaxy and started observing.
The position of the star on its image allowed us to determine
where the radio galaxy should fall on its image with an accuracy
of
about 1 arcsecond.

\bigskip

\centerline{\bf 2.2. Data Reduction}
\bigskip

Since the galaxies we observed were very faint, the procedure
we used to reduce the data, which
used a combination of IRAF routines and our own
programs, was quite elaborate and we
give it here in some detail. It is very similar to
the procedure used by Dunlop \& Peacock (1993), but there are
some
differences.
The first step was to correct for
known bad pixels by replacing the value associated with each
bad pixel by the median of the values in the surrounding
pixels.
The second step was to subtract from
each image of the galaxy an image taken with the same
integration time but with the shutter closed, the effect
of this being
to remove the dark current associated with the
array. The next step was to construct a flat field, by
first normalizing the images of each galaxy so that they had
the same median value, and then
making an image with the intensity at each pixel equal to
the median of the intensities at the same pixel in the normalized images.
After normalizing the flat field so that it had
a mean value of unity, we divided each of the galaxy images
by it.
 
At this point the usual step is to positionally
align the images and then sum them, but we found that a couple
of additional processing steps at this point significantly improved
our images. First, there is the occasional pixel that clearly
has a too high or too low value
compared with the surrounding pixels but which
does not appear in the list of known bad pixels.
We searched for these pixels by looking for
ones that differed by more than a threshold amount from the median
of the surrounding pixels, the threshold being chosen by trial and error
to result in the selection of the obviously bad pixels. We replaced
each bad pixel
by the median of the surrounding pixels. 
Second, on the IRCAM images, even after the flat-fielding step there 
were often large-scale gradients
left on the images. We removed these using the IRAF program IMSURFIT
which fits a polynomial surface to an image and then
subtracts it from the image. The obvious danger in this is that
if the order of the polynomial surface is too high,
some flux will be removed from the objects on the
image. In most cases we avoided this problem completely by only fitting
the polynomial to areas of the array away from the object, and
in practice we found that even when the object was not masked the
step had no systematic effect on the magnitudes but did
result in an improvement to the appearance of the
final image. After
these two additional steps we returned to the usual data-reduction
sequence and positionally aligned the images and summed them. 
In the case of the REDEYE and the IRCAM3 data, we astrometrically
aligned
the images by measuring the position of a
bright object visible in all the images.
In the case of the IRCAM images, with their smaller
field of view, there was often no suitable
object: in such cases we aligned the images
by assuming
that the positioning of the telescope had been accurate. 

Figure
2 shows the final infrared images. 
We have plotted the positions of the radio components on the infrared
images and identified the radio galaxy using two different techniques.
For sources from which optical or infrared emission had not previously
been detected we
identified
the radio galaxy, measured its position, and 
plotted the positions of the radio components
on the infrared image using the astrometric procedure described at
the end of \S 2.1.
There is rarely any uncertainty as to which object
is the radio galaxy, because there is usually either an object close
to the radio core or, where there is no radio core, a single
object midway between the radio lobes. The few cases where there
is some ambiguity are discussed in detail in \S 2.3.
The positions of these galaxies are given in Table 3, and our estimate
of the accuracy of these position measurements is
about an arc second. 
For
sources where the radio
galaxy had already been detected on an optical CCD image, 
we identified the radio
galaxy on the infrared image by comparing the infrared image with the
CCD image. 
In these
cases it is the optical position that appears in Table 3, because
the optical positions are usually more accurate than the positions
we obtained from our IRCAM astrometry.

We measured magnitudes for all the galaxies using the IRAF program
PHOT and circular apertures with a diameter of 5 arcsec and
8 arcsec. In some cases there were two objects close together, and
in these cases we measured magnitudes for both objects. We measured
the sky level by using the IRAF program FITSKY to estimate
the mode of the intensity values in an annulus centered on the
object, with the inner radius of the annulus being made sufficiently
large that the annulus did not include any emission from the
radio galaxy itself.
We set the parameters
of the FITSKY program so that it automatically excluded pixels which
were clearly part of an object, and our use of the mode as the sky
estimator was another precaution against our sky estimates
being biased by the presence of objects in the annulus. 
The magnitudes are listed in Table 4. In cases where there is evidence
that gradients in the sky background are making the large-aperture
magnitudes unreliable we have listed only the small-aperture magnitudes.

The REDEYE observations were made through a $K'$ filter,
but calibrated by observing K-band standard stars (\S 2.1).
Since the $K'$ filter is shifted slightly in wavelength
from the $K$ filter,
and since HZRG's have different colours
from standard stars, there will be a systematic error
in the resultant $K$ magnitudes of the radio galaxies. We estimated
the size of this error, using the transmission functions for the
two filters, an average $H-K$ colour for the radio galaxies
of 1 (Lilly \& Longair 1984), and an average $H-K$ colour for
the standard stars of 0, as 0.13 mag,
in the sense that 
our initial estimates of the K magnitudes had to be made
brighter by this
value. 
Table 4 contains the corrected values.

We have made no correction for galactic extinction, since the high
galactic latitudes of the B2/6C sources mean that the corrections
will be both very small and uncertain. If we use the reddening
map of Burstein and Heiles (1982), we find that two objects have
corrections of about 0.01 mags to their K-magnitudes, with the
corrections for the remaining objects having upper limits of about
this value.

\bigskip

\centerline{\bf 2.3. Notes on the Sources}

\bigskip

\parindent=0pt

{\it 0820$+$36:} The three crosses in Fig. 2 show the positions
of the hotspots and the radio core. There are two objects close
to the radio core, and magnitudes are given for both in Table 4.
The north-eastern object is close to the core, and we conclude that
this is the true identification.

\medskip
{\it 0822$+$34B:} The three crosses in Fig. 2 show the positions
of the hotspots and the radio core. There are two objects close
to the radio core, and magnitudes are given for both in Table 4.
The south-eastern object is closer to the core, and we have assumed
that this is the radio galaxy.
\medskip

{\it 0825$+$34:} There are two objects close to the position of this
small (7 arcsec) double source. There is no central component
to help out with the astrometry, and either of the two objects
might be the radio galaxy. Since the line joining the two objects
is roughly aligned with the radio axis, it is possible that both
objects are components of one object and that this system is an
example of the alignment effect (\S 1). Since the chance of an
unrelated galaxy coinciding with the radio
source is less for a bright galaxy than a faint galaxy, we have
assumed that the brighter of the two objects is the site of the
active nucleus. 
\medskip

{\it 0901$+$35:} There are two objects close to the position of
this small (3 arcsec) double. The northern object is significantly
closer to the position, and we have taken this to be the radio galaxy.

\medskip
{\it 0905$+$39:} We have discussed this source in detail in a separate
paper (Law-Green et al. 1995b).
\medskip

{\it 0918$+$36:} We did not follow our standard astrometric procedure
for this source. There are two objects visible on the infrared
image and both are
also visible on the R-band CCD image of Eales (1985b). Eales 
originally concluded that these objects were not associated with
the radio source because radio galaxies almost always fall within
0.2$\theta$ of the point midway between the radio hotspots
(Laing, Riley \& Longair 1983), where
$\theta$ is the angular distance between the hotspots, and neither
of these objects was within that distance. However, the new
radio map is much better than the original one, and the positions
of the hotspots are slightly different, putting both the galaxies
just within the 0.2$\theta$ circle.
There is no radio core, but the brighter galaxy falls on a direct
line between the hotspots and has the strong emission lines
characteristic of a powerful radio galaxy, 
so it seems 
likely that 
this is the site of the
active nucleus.
The position in Table 3 is a previously unpublished measurement from the 
CCD image.
\medskip

{\it 0943$+$39:} The radio galaxy is unusually far from the point
midway between the hotspots.
\medskip

{\it 1016$+$36:} 
The
radio source is a 15-arcsec double and Allington-Smith suggests
two possible identifications that are visible on his deep CCD image
(Allington-Smith et al. 1982 and
private communication). One is at (1950.0) $\rm 10\ 16\ 58.45$,
$\rm 36\ 37\ 49.7$, and one at $\rm 10\ 16\ 58.59$, $\rm
36\ 37\ 42.4$. There is very little emission present on the
infrared image between the radio
hotspots, with the brightest object visible, the
one marked, having $\rm K = 20.30$. Although it is only
a 3$\sigma$ detection, it is probably real because the position
we measure ($\rm 10\ 16\ 58.48$, $\rm 36\ 37\ 50.6$) is very
close to the position of one of Allington-Smith's objects. 
Although the ratio of the distances of the hotspots to this object
is unusually far from one,
we tentatively conclude that it is the radio galaxy,
but it is also possible that we have not yet detected the true
identification. 
\medskip

{\it 1042$+$39:} We did not follow our standard astrometric procedure
for this source. However, when the UKIRT is pointed at an object
(in this case the point midway between the radio hotspots), the object
usually appears within a few arcsec of the centre of the image.
Since 
there is only one object close to the
centre of the infrared image, we are confident that this is the
true identification. We cannot, however, give an accurate position
for the galaxy.

\medskip
{\it 1045+35A:} There appears to be a cluster of galaxies surrounding
the radio galaxy.

\medskip

{\it 1100$+$35:} The object between the radio hotspots is definitely
the radio galaxy, since it coincides with the radio core. The object
to the west, close to the western hotspot, is probably unrelated,
since its spectrum looks quite different, resembling that of a
low-redshift elliptical (Allington-Smith, private communication).
The large difference between
the K-magnitude
measured by Lilly, Longair \& Allington-Smith (1985) through
a 12-arcsec aperture and the magnitude given here
is probably explained if Lilly et al.'s larger aperture was centred slightly
to the west of the radio galaxy and included the western object.

\medskip

{\it 1143$+$37:} We did not follow our standard astrometric procedure 
for this source. However, the pointing of the UKIRT is usually
good to a few arcsec, and there is only a single object
close to the centre of the image. It is only a 2.6$\sigma$ detection,
but the magnitude is consistent with that given by Lilly (1989).

\medskip
{\it 1204$+$37:} We have assumed that the object closest to the midpoint
of this large (52 arcsec) double is the true identification. 
Evidence supporting this is that we succeeded in measuring a redshift
by centering our slit on the radio midpoint (Rawlings et al. 1996).
\medskip

{\it 1257$+$36:} There are two features on the radio map of Law-Green et
al. (1995a) that may be the radio core. Both are close to
our proposed identification.

\medskip

\centerline{\bf 2.4. Photometric Checks}
\bigskip
\parindent = 20pt

One check we have on our method is that we reduced all the IRCAM data
twice, once using the method described in \S 2.2. and once while still
on Mauna Kea using the standard STARLINK reduction package. Although
the general schemes were similar, the specific programs were different
and there were a number of minor methodological differences: for instance,
in the latter case we estimated the sky level by measuring the mean level
in small apertures that we placed in empty regions of the image. 
There was no systematic trend for one set of magnitudes to be
brighter or fainter than the other.

A second way to investigate the photometric accuracy is to compare
the magnitudes measured with the four different instruments: the three
arrays---IRCAM, IRCAM3, and REDEYE---and the single-element photometer
used by Lilly and collaborators. 
The photometry from the images generally agreed
well with the photometry obtained by Lilly and collaborators using
a single-element photometer (except in the case of
$\rm 1100+35$, where the discrepancy is probably explained by
a nearby object falling in Lilly's aperture---\S 2.3), agreeing with
the conclusion of Dunlop and Peacock (1993) from a similar comparison
for 3C galaxies.
There are four objects where the magnitudes measured
with the HgCdTe camera (REDEYE)
are significantly
different from those measured
with the InSb cameras (IRCAM/IRCAM3): $\rm 0905+39$, $\rm 1011+36$,
$\rm 1017+37$, $\rm 1256+36$. Since the sizes of the differences 
are bigger for the larger aperture, the likely cause is residual
gradients in the sky background, and indeed
on careful inspection of the images it is often possible to
see these gradients.
Residual background gradients appeared to be particular problems
for the IRCAM and REDEYE images, whereas the IRCAM3 images usually
appeared much smoother.
The median discrepancy between the small-aperture IRCAM/IRCAM3
magnitudes and the small-aperture REDEYE magnitudes is 0.09 mags,
in the sense that the REDEYE magnitudes are brighter. However, given
the small number of objects (nine) and
the uncertainty in
the correction from the $K'$ magnitudes to K magnitudes,
we do not believe there
a serious photometric disagreement between the magnitudes measured
with the HgCdTe REDEYE array and those measured with the InSb arrays.

\bigskip
\centerline{\bf 3. The Hubble Diagram}
\bigskip
In this section we compare the infrared Hubble diagrams (K versus z)
of 3C and of the faint samples. 

To construct the 3C Hubble diagram we use two samples of data. Lilly
\& Longair (1984) used a single-element photometer
to obtain K-band photometry of 90 sources
from the 3C sample of Laing, Riley \& Longair (1983) which
(a) were accessible from the UKIRT, (b)
had not been classified by Laing et al. as either a quasar or a probable
quasar, and (c) did not have redshifts $\rm < 0.03$.
There are an additional seven sources in the sample
of Laing et al. satisfying these criteria which were not
observed, four because
they are close to bright stars, making reliable photometry
difficult, and three because the deep optical images necessary
for the success of Lilly and Longair's observational technique
were not available (in one case because of the presence of a bright star,
in the other two because the sources were late additions to the 3C sample).
We are confident that the omission of these seven sources should not
have biased Lilly and Longair's results and should not bias ours.
In two cases
the optical identification
assumed by Lilly \& Longair is incorrect (3C13 and 3C326---Le
F\`evre et
al. 1988; Rawlings et al. 1990),
with the result
that the K magnitude of the wrong galaxy was measured. We have discarded the
data for these galaxies.
The other sample is that of Dunlop \& Peacock (1993), who obtained K-band
images of 19 radio galaxies in the redshift
range $\rm 0.8 < z < 1.3$ from the larger 3C sample listed
by Spinrad et al. (1985). There are other 3C galaxies in this redshift
range, but Dunlop and Peacock claim that the sample of galaxies they did
observe is a representative sample of the galaxies in this redshift
range.
We are only interested in objects
for which there is not already evidence that the optical/infrared
light is mostly nonstellar, and so we
have excluded from our analysis all radiogalaxies that Spinrad et al. (1985)
classify as N-galaxies and ones that have the broad permitted emission-lines
typical of quasars. 

Many of the radio galaxies fall in both samples, and there is then
the question of which magnitude to use. 
Dunlop \& Peacock
(1993) have shown that their magnitudes agree well with
Lilly \& Longair's magnitudes. 
We decided to use an array magnitude when one exists, 
simply because one is then
sure about the position of the aperture and that it does
not include the flux from some unrelated object.
The redshifts for the 3C galaxies are mostly from Laing et al.
or from Spinrad et al. (1985), with the exception
of a few unpublished redshifts mostly communicated to us by Spinrad
(from whom an up-to-date list of 3C redshifts can be obtained).

For the faint sample there are several
sources of infrared magnitudes: (a) the magnitudes
listed in this paper, (b) IRCAM magnitudes given in Eales et al. (1993a, 1993b)
and in Eales \& Rawlings (1996), (c) magnitudes measured with the UKIRT
single-element photometer by Lilly, Longair \& Allington-Smith
(1985) and by Lilly (1989). With the exception of the REDEYE magnitudes
given in this paper, all the magnitudes have been measured with either the same
instruments (or a very closely-related one, IRCAM3) 
used to measure the magnitudes of the 3C galaxies. 
The data reduction procedures have also been either the same or
very similar. As for the 3C galaxies, if an array magnitude exists we used
it. If more than one array magnitude exists, we used 
an IRCAM3 magnitude,
because of the flatness of the IRCAM3 images, when possible, and when not
the average of the REDEYE and IRCAM magnitudes, it being unclear which
of these is preferable.
For the galaxies for which there is a large difference between
the IRCAM and the REDEYE magnitudes, we used the average of the magnitudes
measured through the smaller aperture, for which the magnitude difference
is always less (\S 2.4).
Note that the photometric agreement
between the different datasets (\S 2.4) is sufficiently good, that
our choice of magnitudes should have a negligible effect on our results.
The redshifts for the 6C/B2 galaxies that have not yet been published
will be listed in Paper II. As we are only interested in objects
for which there is not already evidence that the optical/infrared
light is mostly nonstellar, we
have again excluded from our analysis all the objects with the broad permitted emission-lines
typical of quasars.

Now let us consider how to estimate magnitudes that correspond to the
same metric aperture at all redshifts. Ideally, one choses a 
metric aperture, and
then observes each galaxy through the aperture that gives
this physical size at the redshift of the galaxy. In practice, this
is not entirely possible, although one can approach this ideal
by a careful choice of the metric aperture. The universe
helps out to some extent, because above a redshift of 0.5
the physical distance subtended by a given angle depends
only weakly on redshift. Since most of our photometry was obtained
through an aperture of diameter eight arcsec or close to it,
we have chosen 63.9
kpc as our metric diameter, since for $\rm \Omega_0 = 1$ this
corresponds to $\simeq$8$''$ at a redshift of one.
Thus
much of our photometry was obtained through an aperture very similar
to our chosen metric aperture. For most galaxies, though, some 
correction, however small,
has still to be made to the measured magnitudes to convert them
into magnitudes measured through the metric aperture. The standard
way to do this is to assume
that all radio galaxies have the same intensity profile, and
then use this standard profile to correct the magnitudes. This is a very
good approximation at low redshift, where most radio galaxies are
giant ellipticals, and the intensity profile or ``curve of growth''
tabulated by Sandage (1972) is the one that has traditionally been
used (e.g. Laing, Riley \& Longair 1983). This is clearly the wrong
thing to do here, because the structures of HZRG's
often look very unlike those of low-redshift giant ellipticals
(Fig. 2 and Dunlop \& Peacock 1993). We decided to adopt the following
compromise, which, on the one hand, avoids assuming that the structures
of radio galaxies do not evolve and, on the other hand, avoids the 
necessity of estimating curves of growth for each individual
galaxy. We divided the galaxies into those below a redshift of 0.6
and those above this redshift, because this is the redshift above which
the alignment effect is seen (McCarthy 1993). Below this redshift it
seems a reasonable approximation that radio galaxies do have growth
curves like that of a giant elliptical, and we used Sandage's growth
curve to correct our magnitudes to the standard metric aperture. Above
this redshift we assumed that the emission within an aperture of radius
$r$ is $\propto r^{\alpha}$ and derived a value for $\alpha$ of 0.35
from the median difference
between the small-aperture and large-aperture magnitudes in Table 4
(0.18 mags). We then used this intensity profile to correct all
the magnitudes of galaxies at $\rm z > 0.6$ to the standard metric
aperture. It is important to note that the aperture corrections
are generally small. At $\rm z > 0.6$ 75\% of the galaxies have aperture
corrections less than 0.05 mags.

Figure 3 shows the Hubble diagram for the two samples. There are some
obvious
visual differences between two distributions. We have adopted the following 
method for analysing those differences. There are two obvious redshifts
at which to split the diagram. First, there are no 3C galaxies at $\rm z > 1.8$
and so we have removed all the 6C/B2 galaxies above this redshift from
our analysis. Second, the alignment effect starts to be seen at redshifts
$\rm >0.6$ (McCarthy 1993), so this seems a natural redshift to split the
data. In the two redshift bins---$\rm 0 < z < 0.6$ 
and $\rm 0.6 < z < 1.8$--- we have used the following method of seeing
whether the distributions are similar for the two samples. In 
each bin we fitted
a straight line to the data points for 3C by minimizing the sum of
the squared residuals
in magnitude between the points and the line. We then plotted histograms
of the residuals between the data points and the line for both the
3C and the 6C/B2 sample. The fitted lines are shown in Figure 4, the
histograms in Figure 5.

Figures 5(a) and 5(b) show the residuals for the low-redshift bin. There
is no obvious differences between the histogram for 3C and that
for the faint sample, and this is borne out by a Mann-Whitney U-test,
which shows that the null hypothesis that the two samples are drawn
from the same underlying distribution is statistically acceptable.
This agrees with the results of a study by Hill and Lilly (1991)
of a sample of radio galaxies at $\rm z \sim 0.5$. They showed
that for this sample the absolute optical magnitude is independent
of radio luminosity over a range of $\simeq$1000 in radio luminosity.
Thus our result and that of Hill and Lilly agree with the
conventional wisdom that low-redshift radio galaxies are `standard
candles' with a small range of absolute magnitude, and
that the absolute magnitude is independent of radio luminosity.

Figures 5(c) and 5(d) show the residuals for the high-redshift sample. 
Here the distributions are clearly different, and application of the
Mann-Whitney U-test shows that the probability
that the null hypothesis, that the distributions
are drawn from the same underlying distribution, is correct 
is $\simeq$0.01\%.
This is also very clear from Fig. 4. Almost all the B2/6C galaxies
are fainter than the line that is the best fit to the 3C data. The
median difference between the 3C and 6C/B2 residuals is 0.59 mags (a factor
of 1.7 in intensity)\footnote\ddag{This is slightly less than the 
difference claimed in Eales \& Rawlings (1996), the change being due
to remeasurements of some of the redshifts.}.
Since the only other difference between the 3C and B2/6C galaxies
is one of radio luminosity (roughly a factor of 6), in this redshift
bin there is a correlation between
radio and infrared
luminosity\footnote\dag{Because of the redshift, the emission we are detecting
in the infrared K-band is being emitted by the galaxies at 
$\rm \sim 1 \mu m$, which is actually on the border between the infrared
and optical wavebands.}. 

We cannot think of any obvious way our analysis could have spuriouly
generated this result. Although there are some differences between the
magnitudes measured with different cameras (\S 2.4), the differences
are much smaller than the size of the average difference between
the magnitudes of the 6C/B2 and 3C galaxies. It would be possible
to produce this result if we had systematically misclassified more
3C quasars as galaxies than 6C/B2 quasars as galaxies. If anything, though,
we have probably erred in the opposite direction, since we excluded
3C sources classified in the literature as N-galaxies, whereas  
no such classifications exist for the 6C/B2 galaxies. 
Finally we consider the eight 6C/B2 galaxies that are not plotted
in the Hubble diagram because they do not yet have redshifts. One of these
we can ignore because the lack of a redshift is simply due to the presence
of a nearby star, and thus its omission from the diagram does not
bias the diagram in any way; two are relatively bright galaxies ($R \sim
20$), and have estimated redshifts which put them in the low-redshift
bin; one is a BL Lac object; 
the remaining four might be galaxies in the high-redshift bin, and so, in
principle, their omission might affect our result. The K-magnitudes
of these four are 19.8, 19.0, 17.36, and 19.46. Three of these are
among the faintest K-magnitudes that we have measured, and if these
objects do fall in the high-redshift bin the true difference between the
B2/6C and 3C galaxies will actually be greater.
Therefore, it
seems likely that the difference between the K-magnitudes of the B2/6C and
3C galaxies is a fundamental one.

It is possible to explain all the properties of the Hubble
diagram---the result that
there is no difference between the samples at low redshift, that
there is a difference at high redshift, 
and Lilly and Longair's discovery 
that the infrared luminosities
of 3C galaxies increase with redshift---by the following simple model.
We will assume that there are two components to the near-infrared emission
from a radio galaxy: starlight and nonstellar light. We will
not worry about the nature of the second component until
later (\S 5.1); for now we
will assume that its strength is proportional to the power
output of the active nucleus. Since the radio luminosity will also
monotonically increase with the power output of the active nucleus
(Rawlings \& Saunders 1991), the nonstellar infrared component
will be proportional to the radio luminosity. We will additionally assume
that the starlight component does not depend on radio luminosity. 
There is no guaruntee that this model is correct, of course, but it
is the least radical departure from our existing ideas about radio galaxies
that we could think of; it keeps the idea that the galaxies themselves
have a small range of absolute magnitude at a given redshift
while incorporating the idea (for which there is evidence
even at low redshift---Yee \& Oke 1978) that there is a nonstellar component
to the light from radio galaxies. If we assume this model, the properties
of the Hubble diagram follow quite naturally. It explains why at low redshift
there is no difference between the B2/6C
and 3C galaxies: the radio luminosities of both sets
of galaxies are then so low that the nonstellar infrared emission is swamped
by the starlight from the galaxy. 
It explains why the 3C
galaxies at $\rm 1 < z < 2$ have higher infrared luminosities than
the B2/6C galaxies in the same redshift range: the 3C galaxies have higher
radio luminosities and thus more nonstellar emission. Finally it explains
why 3C galaxies at $\rm z \sim 1$ are more luminous than those
at $\rm z \sim 0$: because of the correlation between radio luminosity
and redshift in any flux-limited sample, 3C galaxies at high redshift
have higher radio luminosities, and thus higher
infrared luminosities, than those at low redshift. 
In contrast, Lilly
\& Longair's model cannot explain all the properties
of the Hubble diagram. If stellar evolution is responsible for
the increase in infrared luminosity of the 3C galaxies between now and
$\rm z \sim 1$, it should have a similar effect on the 6C/B2 galaxies,
since the optical properties of the low-redshift 6C/B2 galaxies are
indistinguishable from those of the low-redshift 3C galaxies.
\bigskip
\centerline{\bf 4. The Alignment Effect in the Infrared}
\bigskip
Since there is no agreement about the explanation of the alignment
effect, any information about how it
depends on redshift,
radio luminosity, and on wavelength of observation is important.
Rigler et al. (1992), in a study of 3C galaxies at $\rm z \sim 1$,
found that, in contrast to the optical structures,
there is no evidence that the infrared structures are aligned with
the radio structures. They also found that the infrared
structures are generally less elongated than the optical structures.
In contrast, Dunlop \& Peacock (1993; henceforth
DP), in a study of a very similar sample
of 3C galaxies, found that the infrared structures of 3C galaxies
at $\rm z \sim 1$ are actually more closely aligned with the radio
structures than the optical structures are, although they agreed with
the conclusion of Rigler et al. that the infrared structures are generally
less elongated. They extended their study of the alignment effect to galaxies
with lower radio luminosities by selecting
galaxies from a faint radio sample drawn from the
Parkes radio catalogue  with a similar range of K magnitude
to their sample of 3C galaxies. As long as the relation between K magnitude
and redshift is the same for both samples, the Parkes galaxies will lie
in the same redshift range as the 3C galaxies, but will have lower radio
luminosities. DP
found that although the infrared structures of the
3C galaxies are closely aligned with the radio structures, the infrared
structures of the Parkes galaxies are not, implying that the alignment effect
is a function of radio luminosity.

We argued in the last section that the basic assumption of DP
is not true: the relation between K magnitude and redshift is 
not the same for a bright and a faint radio sample, the relation for
the faint sample being offset to fainter magnitudes. The effect of this will
be that galaxies in DP's faint radio sample will actually 
lie at
redshifts that are lower than the ones predicted from the 3C relation.
Thus their two samples of HZRG's---the 3C sample
and the Parkes sample---will not cover identical redshift ranges; the
Parkes galaxies will generally be at lower redshift. We do not claim their
result is invalid, because the main difference between
the two samples, and thus the likely explanation of the difference in the
alignment effect, is still likely to be one of radio luminosity rather
than redshift, but this effect does produce some uncertainty in their
result. However, we can reexamine the question, and hopefully dispel some
of the uncertainty, because we can use redshifts rather than K magnitudes
to compile a suitable sample of faint radio galaxies to compare with 
DP's 3C sample.

DP considered the redshift range $\rm 0.8 < z < 1.5$, and
there are 15 B2/6C galaxies in this redshift range. Of these
15, there are nine which have
both the FR2 radio structure 
(Fanaroff \& Riley 1974) and for which there are infrared
images.
Because estimates by eye of the position angle of an infrared
structure are uncertain, depending on such things as the
greyscale used to display the image, DP developed an objective
method of estimating the position angle.
They calculated
the intensity moments ($<Ixy>$, $<Ix^2>$, $<Iy^2>$) of each galaxy,
using only pixels within an angle $\theta$ of the centre of the
galaxy,
and then they diagonalized the matrix of moments to find the position angle. 
To maximize the signal-to-noise of their estimates, they only considered
pixels whose intensities were greater than $0.3I_{max}$, $I_{max}$ being
the maximum intensity of each galaxy. Their choice of $\theta$ was governed
by the size of the radio source. For sources with angular sizes greater
than 8 arcsec they used a value for $\theta$ of 4 arcsec;
for sources with a diameter of less than 5 arcsec they a value for
$\theta$ of 2.5 arcsec;
and for sources with a diameter between 5 and 8 arcsec
they used a value for $\theta$ equal to half the diameter
of the source (see
DP for a justification of this). We used exactly the same technique as
DP. For galaxies where we had more than one image, if possible
we estimated the
position angle from a REDEYE image, because these had the best angular
resolution (FWHM usually less than one arcsec), and if a REDEYE image was
not available we preferred the IRCAM3 image to the IRCAM image. 

Our estimates of the position angle are given in Table 5, and a histogram
of the differences between the infrared and radio position angle is
shown in Figure 6. We used a Kolmogorov-Smirnov one-sample test (Siegel
1956)
to examine whether the data are consistent with a uniform distribution
(i.e. no alignment effect). The result was tantalising, since we found that
the chance that this null hypothesis is correct is $<$20\%; the limit
arising because the standard statistical tables (Siegel 1956) consider
the probability that the null hypothesis is true against all possible
alternatives, whereas we are only interested in the alternative
hypothesis that the data is bunched up towards 0$^{\circ}$---and the
data is in this direction. Nevertheless, in contrast to the
histogram for 3C (DP), there is no evidence of a very strong effect,
supporting DP's conclusion that the alignment effect
is a phenomenom associated mostly with the most luminous sources.
\bigskip
\centerline{\bf 5. Discussion}
\bigskip

\centerline{\bf 5.1. The Nature of the K-band Excess}

\bigskip

In \S 3 we sketched a possible explanation of the various features
of the Hubble diagram by suggesting  there are two 
components to the near-infrared
emission from a radio galaxy: light from stars and a nonstellar component
that is produced by the active nucleus and
whose strength is correlated with the total radio luminosity. We 
claimed that this is the least
radical change one can make to our existing ideas about radio galaxies and
still explain the Hubble diagram. However, at that point we did not fill
in any of the details of the sketch. There are three
ways one can do this.

One possibility is that some of the K-band emission is actually due to
emission lines rather than to continuum processes. 
It has been known for a while that
emission-line luminosity is correlated with either radio
luminosity or redshift (Rawlings et al. 1989), and a comparison of 
emission-line fluxes for the B2/6C galaxies with those for 3C shows
that the true correlation is with radio luminosity (Rawlings et al.
1996). So if the K-band flux is being produced by emission lines
in the K-band filter, one would
expect a 
correlation between K-band luminosity and radio luminosity. However, at
these redshifts the potentially troublesome lines are [SIII] 953.2
and He I 1083.0, which have lower equivalent widths than [OIII] 500.7
and H$\alpha$, which significantly ``pollute'' the K-band magnitudes
of HZRG's at $\rm z > 2$ (Eales \& Rawlings 1993, 1996; Eales et
al. 1993b). The infrared spectroscopy that we have done of 3C galaxies
at $\rm z \sim 1$
(Rawlings, Lacy \& Eales
1991, and unpublished data)
shows that the contribution of emission lines is {\it at most}
25\% of the total K-band emission. Therefore it seems unlikely
that the infrared-radio correlation is caused by line emission.

An alternative is to propose that there are quasars in the centres
of the radio galaxies, that the optical/infrared continuum emission
from these quasars is proportional to the radio luminosity, and
that these quasars are behind screens of dust
which are just thick enough to conceal them from our view at optical
wavelengths but not at infrared wavelengths. 
If this model is correct,
a large fraction of the near-infrared light
from a high-redshift 3C galaxy is actually coming directly from a quasar.
There must be enough dust
present for the broad permitted lines and bright 
continuum emission that are characteristic of a quasar to be hidden
at optical wavelengths, because
otherwise the object would already have been classified as a quasar.
There are a number of things in favour of this model. 
First, it is broadly in line with the popular ``unified models''
(e.g. Barthel 1989).
Second, Serjeant et al.
(1996) have shown that for quasars with steep radio spectra---that is
quasars with similar radio properties to most 3C radio galaxies---radio
and optical luminosity are correlated. This does not of course prove that
the model is right, merely that if there are quasars in the centres
of the radio galaxies obscured by a small screen of dust then there
is reason to expect the near-infrared continuum emission to scale with
the radio luminosity. Finally, we have shown that for one object this
model does seem to be correct. The radio source 3C 22, which has a redshift
of 0.937, is classified as a radio galaxy, because it has
narrow permitted optical emission lines. However,
we have recently observed this with an
infrared spectrometer and found that H$\alpha$, which is redshifted into
the near-infrared J-band, is broad (Rawlings et al. 1995). 
The lack of broad permitted lines in the optical spectrum
and the red spectral energy distribution can be explained if we are
viewing a quasar through a screen
of dust with a visual extinction of $\sim$ 2 mags. 

Nevertheless, although there is some evidence in favour of this model, it
is not entirely satisfactory. First, it does not explain why there is a
near-infrared alignment effect for high-redshift 3C galaxies but not
for high-redshift B2/6C/Parkes galaxies (\S 4). If anything one would
expect the reverse. If the light from a high-redshift 3C galaxy
is dominated by light directly from the central quasar, then this light
would make the alignment effect, which must be connected to the other
component of the light, harder to see. Second, there are reasons to
think 3C 22 might be different from the other high-redshift 3C galaxies.
It has a one-sided radio jet (Fernini et al. 1993), it has by 
far the brightest K-magnitude of any of the high-redshift
3C galaxies, and of the objects observed by DP it is the one whose
K-band image looks most like that of a quasar; none of the infrared structures
of the other 3C and Parkes galaxies is so clearly dominated by a single
unresolved source.

It is possible to avoid these problems and retain the general outline of
the model if, instead of the quasar light being seen directly, it is
scattered towards us by either dust or electrons. There is evidence
that this process operates at optical wavelengths. The degree of 
polarization of the optical light from high-redshift 3C galaxies
is typically around 10\% and the electric vector of the polarized
light is usually perpendicular to the radio axis (e.g.
Tadhunter et al. 1992; di Serego Alighieri, Cimatti \& Fosbury 1993), exactly
what is expected if the central quasar in a radio galaxy is completely
hidden
from our direct view but is illuminating dust or electrons along the radio
axis (in the unified scheme of Barthel [1989]
assumed to be close to the plane of the sky), which scatter
the quasar light towards us.
This scheme would both explain the features of the Hubble diagram and
explain why the alignment effect is seen clearly for the 3C galaxies but not
for the 6C/B2/Parkes galaxies.
An obvious argument against this scheme is that the spectral energy
distributions of HZRG's increase to longer wavelengths,
whereas the scattering efficiency of electrons remains constant, and that
of dust decreases, towards longer wavelengths. However, it is relatively
easy to think of a model that could produce more light being scattered
in the infrared than at optical wavelengths if one allows for the
absorption produced by dust. If there is a fairly large amount of dust
along the radio axis, the scattering particles, whatever they are, may
see little optical emission, and although
the scattering efficiency may be less
at longer wavelengths there may be much more light to scatter. There is some
limited evidence that in the optical waveband, the degree of polarization
decreases to longer wavelengths 
(Cimatti et al. 1993), but the evidence 
is not 
compelling and, even if correct, it would be still be possible for
the degree of polarization to start increasing again. The real way
to test this model is to perform imaging polarimetry in the near
infrared, and we are in the process of doing this. 

We argued that this overall model is the least radical departure
from our existing ideas, but this does not mean that it is correct.
It is quite possible that the true explanation is completely different.
We will just mention one idea as an example of more exotic explanations.
Even at a redshift of one, galaxies with radio luminosities as high as those
of 3C galaxies are quite rare, with there being $\sim 10^2$ such objects
over the whole sky. Our current ideas about the physics of double
radio sources suggests there are two ways to increase the low-frequency
radio luminosity of a galaxy (see Rawlings
\& Saunders 1991; Eales 1992). One is to increase the energy flowing
along the jet, roughly the power of the central engine, and one
is to increase of the density of the gas surrounding the radio lobes.
Suppose that to generate the radio luminosity of these rare ultraluminous
3C galaxies both factors are necessary: the most powerful central engines
and also the densest gas. We know very little about galaxies at a redshift
of one, but it seems quite likely that the most extensive and densest
distributions of gas are likely to be associated with the most massive
stellar systems. Thus, if this idea is correct, the 3C galaxies
at $\rm z \sim 1$ are the most massive systems around at this epoch
 and the K-band light
is still starlight. Of course this is only a sketch (a similar and
more detailed sketch 
is given in Eales, 1992), but it seems to us
that this is as likely to be true as the less exotic sketch shown
above.
\bigskip
\centerline{\bf 5.2. Galaxy Evolution}
\bigskip
Figure 7a shows the Hubble diagram and the predictions of
various models. 
The broken lines show the predictions of various evolutionary
models. We have produced these using the models of
Bruzual \& Charlot (1993), which are
contained in the ``Galaxy Isochrone Synthesis Spectral Evolution Library''
(GISSEL). These models allow the user to investigate the
ultraviolet/optical/infrared properties of a galaxy for any
history of star formation and for a number of different initial
mass functions. The specific GISSEL model we have used
is the one for an instantaneous burst of star formation
with a
Salpeter IMF with a lower mass cutoff
of 0.1 M$_{\odot}$ and an upper mass cutoff of 125 M$_{\odot}$.
The three models correspond to three different assumptions about
the history of star formation in a radio galaxy, although in each case
the star formation is assumed to occur at $\rm z >> 1$.
The continuous line is the prediction of a no-evolution
model\footnote\dag{All that is necessary to calculate this is the
average spectral energy distribution of
low-redshift radio galaxies. A minor problem is that although
there are published spectral energy distributions of adequate
spectral resolution in the optical waveband (e.g. Coleman, Wu \&
Weedman 1980), the only spectral information in the near infrared is
broad-band colours. We have overcome this by using the GISSEL models
to find a star-formation
history for a radio galaxy that produces, at the current
epoch, both the observed near-infrared
colours and the observed optical spectral energy distribution. The models
can then be used to generate a zero-redshift spectral energy distribution
in the near infrared of sufficient spectral resolution
to generate a K-z curve. All that we are really
doing is finding a physically plausible way of
interpolating between the near infrared magnitudes of a low-redshift
radio galaxy, and although
in principle this does not produce a unique solution, with a variety
of star-formation histories capable of generating the same 
observed optical and near-infrared properties at low redshift, in 
practice we found that different star-formation histories, as long as they
could reproduce the observed low-redshift properties, lead to virtually
identical K-z curves.}. We have normalized all the models so that their
predictions
pass through the low-redshift points.  
As Lilly \& Longair (1984) discovered, the 3C
galaxies at $\rm z \sim 1$ are significantly brighter than the no-evolution
curve, and agree rather well with the predictions of the evolution models.

However, we have concluded that the bright K-magnitudes of 3C galaxies
at $\rm z \sim 1$ are connected to the high radio luminosities of these
objects, not to their redshifts, and so the agreement between the magnitudes
and the evolutionary models is actually fortuitous. Is it still possible
to draw any conclusions about galaxy evolution from this diagram?  
It turns out that it {\it is} possible to draw some conclusions, although only
tentative ones, and ones which depend on the assumption one makes
about the explanation of 
the relationship between infrared and radio luminosity.

First, let us assume that our conservative hypothesis is correct, that
there are two components of the near-infrared light: 
light from stars and a nonstellar component
that is produced by the active nucleus and
whose strength is correlated with the total radio luminosity.
At any redshift, the nonstellar component will be less important
for the 6C/B2 galaxies than for the 3C galaxies, and so
inferences about galaxy evolution drawn from the 6C/B2 K-z relation
are more likely to be correct
than those drawn from the 3C relation. 
In Figure 7a,
the 
B2/6C galaxies actually follow
the no-evolution curve out to a redshift close to two. 
This is rather a surprise, because stellar evolution should
be making the galaxies brighter in the past. However, it is a phenomenom
that is also seen in the K-z relation for galaxies found in deep
surveys. Although the K-z relation for normal galaxies does
not have the small dispersion seen for radio galaxies, and so it
is not useful to try and fit a curve to the points, it is useful
to look at the upper envelope of the distribution---the line that
traces the brightest K-magnitude seen for normal galaxies at any redshift.
Songaila et al. (1995) have found that, out to $\rm z \sim 1$,
this line also does not
show the brightening expected
from stellar evolution. Both for our radio galaxies and the
normal galaxies, there are two obvious possible solutions to
this paradox. First, we may have assumed the wrong cosmology. For
a low-density Friedmann universe (Fig. 7b), the 6C/B2 galaxies are
brighter than the no-evolution model at $\rm z \sim 1$. Second, there
may be another type of evolution occurring that is operating
in the opposite direction to that of stellar evolution. 
All models of galaxy formation based
on hierarchical clustering (Press \& Schechter 1974; White et al.
1983) make the firm prediction that the structures
of galaxies should evolve at the same time as the stars in them
evolve.
Indeed, simulations
based on these models produce images that look
remarkably like the linear lumpy images of HZRG's (e.g. Fig. 3,
Baron \& White 1987). Since this process will make
the luminosities of galaxies increase with time, it would
tend to cancel out the efffect of stellar evolution, and thus
might explain the overall lack of evolution seen in the Hubble
diagram.

These two possible conclusions depend, of course, on whether the
basic hypothesis is correct. They also depend on one other assumption.
We have implicitly assumed that although the near-infrared light
from 3C galaxies at $\rm z \sim 1$ is dominated by nonstellar light,
the near-infrared light from the high-redshift 6C/B2 galaxies is
mostly from stars, whereas it is possible
that this is still ``polluted'' by nonstellar light. If that is so,
the magnitudes of the {\it stellar} components of the 6C/B2 galaxies would
be displaced upwards in the Hubble diagram from the existing points,
strengthening the evidence for merging if $\rm \Omega_0 = 1$, and
possibly requiring it for $\rm \Omega_0 = 0$ as well. This, however, does not
seem too likely because optical spectroscopy of 3C galaxies has found
evidence of a significant stellar component even in these objects
(Lacy \& Rawlings 1994; Lacy et al. 1995; Stockton, Kellog 
\& Ridgeway 1995).
The obvious way to test this idea is to look at the Hubble relation of
an even fainter radio sample, where the luminosity of the nonstellar
component should be even less. We and other groups are in the process
of doing this. As yet, there is a single intriguing result.
Dunlop et al.
(1996) have obtained a deep spectrum with the Keck Telescope of a
radio galaxy at a redshift of 1.55. The spectrum is very similar to
the rest-frame $UV$ spectrum of a low-mass 
main-sequence star, implying that it is dominated by stars,
and the galaxy's radio luminosity is many times less than that of 6C/B2
galaxies at the same redshift. Nevertheless, it 
is in a similar place in the Hubble diagram to the B2/6C galaxies
(Figs 7), suggesting that the light from the B2/6C galaxies
is dominated by stars as well, and that, after discarding the polluted
3C galaxies, the old assumption that radio galaxies
are ``standard candles'' is
valid out to a $\rm z \sim 1-2$.

As we noted in the previous section, merely because an assumption is
the most conservative one does not necessarily make it correct; and
if we start out with a different explanation of the relation between
radio and infrared luminosity we end up with a different conclusion about
galaxy evolution. As an example of more exotic explanations, let us consider
again the suggestion in \S 5.1 that, at $\rm z \sim 1$, the most radio-luminous
sources (those in 3C) can only exist in the most massive galaxies, 
whereas slightly less radio-luminous sources (6C/B2) can survive in less
massive galaxies. At the current epoch radio galaxies with a wide range
of radio luminosity
are always associated with the most massive galaxies, so this hypothesis
leads to an opposite conclusion to the hypothesis above. It is no
longer correct to compare high-redshift 6C/B2 galaxies with low-redshift
radio galaxies, because one is then comparing apples and oranges, or
rather the most massive galaxies at the current epoch with galaxies
further down the galaxy luminosity function at $\rm z \sim 1$. Instead,
it is reasonable to compare the 3C galaxies at $\rm z \sim 1$ with low-redshift
radio galaxies, because one is then comparing the most massive galaxies at 
both epochs. One then of course reaches precisely the same conclusion
as Lilly \& Longair (1984): the most massive galaxies are brighter at
$\rm z \sim 1$ than they are today, possibly because of stellar evolution.
We have now discussed the consequences for
galaxy evolution of two possible explanations 
of the relation between
radio and infrared luminosity, but there are undoubtedly other
possible explanations. 
Thus, to summarize, one can draw conclusions 
about galaxy evolution, but they depend on the assumptions one makes about
the explanation of the relation between
radio and infrared luminosity.

The discussion above applies only to redshifts less than $\rm \sim 2$,
because, as Fig. 7 shows, above this redshift the galaxies are 
systematically brighter than the no-evolution model and the
dispersion in the K-z relation also appears to
increase.
We have already suggested
this second result
in a previous paper (Eales
et al. 1993a), in which we looked at the 
K-magnitudes of the handful of radio galaxies that were then known
to have redshifts greater than two and which had K-magnitudes. This 
sample
did not form
a statistically-complete sample and for this reason
our conclusion was very tentative. Here we can be much more confident, 
because we are comparing flux-limited samples of sources for almost
all of which we have
redshifts and K-magnitudes.
Let us consider the data in two natural
redshift bins: the 3C and the 6C/B2 galaxies together at $\rm z < 0.6$,
and the 6C/B2 galaxies at $\rm z > 1.8$, the maximum redshift in 3C.
As in the earlier paper, we fit straight lines to the points in both
samples, minimizing the sum of the root mean squared residuals in magnitude. 
The dispersion around the line for the low-redshift points is 0.47 mags;
for the high-redshift bin it is 1.29 mags. 
This is a larger difference than we found in the previous paper, it is
for a larger and more uniform sample of HZRG's, and the
result is thus much more secure. Although we are missing redshifts
for four faint galaxies, it is easy to see by inspection of the
Hubble diagram that including these in the high-redshift bin
would not significantly change this result.

Both of these changes in the K-z relation at $\rm z \sim 2$ are suggestive
of galaxy formation. If the stars in a galaxy form at a constant rate
in a specified period of time, the maximum optical luminosity of the
galaxy occurs at the end of this period (Charlot \& Bruzual 1991). The
minima in the evolutionary curves shown in Figure 7a occur at this
time, and several of the 6C/B2 galaxies have magnitudes very similar
to the ones predicted for the end of this formation period. An
increase in dispersion is also a predicted feature of the epoch of
galaxy formation. This follows because the luminosity of a galaxy
will change most rapidly immediately after
the period in which most of the stars are formed
(e.g. Figure 5 of Charlot \& Bruzual 1991). Therefore, as long as
radio galaxies do
not form at exactly the same instant, the range of their luminosities should
be greatest during the formation epoch. However, although both of these
properties of the K-z relation do suggest that we are seeing the 6C/B2 galaxies
at $\rm z > 2$ either shortly after or during the epoch in which most
of the stars in the galaxies formed, there are two factors that have to 
be considered. First, at $\rm z > 2$ there are potentially bright emission
lines that fall in the K-band and may inflate the K-band fluxes. Second,
the 6C/B2 galaxies at $\rm z > 2$ have higher radio luminosities than
those at lower redshift, and it is possible that the kind of selection
effect we have argued is responsible for the bright magnitudes of the
3C galaxies at $\rm z \sim 1$ is also responsible for the bright
magnitudes of the 6C/B2 galaxies at $\rm z > 2$. We have addressed all of these
issues in a second paper (Eales \& Rawlings 1996).
\bigskip
\centerline{\bf 6. Summary}
\bigskip

We have obtained redshifts and infrared images for a sample of faint
B2/6C radio sources whose fluxes are about six times fainter
than those of sources in the bright 3C sample. 
By comparing the infrared structures
and Hubble relations of the bright and faint samples, we have reached
three conclusions:
\parindent = 0pt
\medskip
(1) At low redshift radio galaxies are ``standard candles'', with
radio galaxies almost always being giant ellipticals with a small
spread of absolute magnitude.
However,
at $\rm z \sim 1$ we find that the 6C/B2 galaxies are on average 
$\simeq$0.6
mags
fainter in the K-band than 3C galaxies at the same redshift.
The 6C/B2 galaxies have radio luminosities which are about six times
fainter than those of 3C galaxies at the same redshift, and thus
at high redshift radio and infrared luminosity are correlated.
\medskip
(2) The infrared stuctures of
B2/6C galaxies at $\rm z \sim 1$ are less clearly
aligned with their radio structures than 3C galaxies at
this redshift, supporting
the conclusion of Dunlop \& Peacock (1993) that the strength of
the alignment effect depends on radio luminosity.
\medskip

(3) Out to a redshift of $\rm \sim 2$ the 6C/B2 galaxies follow
a no-evolution model ($\rm \Omega = 1$), but above this
redshift the galaxies are systematically brighter than this
model and the
dispersion in the K-z relation 
is
about 2.7 times that at low redshift.
\medskip

\parindent = 20pt

The most conservative (in terms of changes to our present ideas) explanation
of the first result is
if there are two
components to the near-infrared
emission from a radio galaxy: light from stars and a nonstellar component
that is produced by the active nucleus and
whose strength is correlated with the total radio luminosity.
If the nonstellar light is coming directly from the active nucleus
(seen through a screen of dust to hide, at optical wavelengths,
the signs of a quasar),
the model cannot explain the second result; but if the
nonstellar light is being seen indirectly, scattered towards us by
either dust or electrons, the model can explain both the first and
second result. If either of these ideas is correct, Lilly \& Longair's conclusion that the
the bright K-band magnitudes of 3C galaxies at $\rm z \sim 1$ are due
to stellar evolution is incorrect;
they are actually due to the high radio luminosities of these objects.

In this scenario, since the infrared magnitudes of the 6C/B2 galaxies 
will be less ``polluted'' by nonstellar emission, it may still be possible
to make inferences about galactic evolution by looking at the 6C/B2 
K-z diagram. Out to $\rm z \sim 2$, these galaxies follow the 
no-evolution curve, 
in contradiction to the prediction
that stellar evolution should be making galaxies brighter in the past.
Possible ways out of this paradox are if we live in a low-density
universe or if merging is making galaxies fainter in the past. However,
we emphasize that inferences about galactic evolution depend critically
on the assumptions one makes about the cause of the correlation between
infrared and radio luminosity, and it is possible to think
of qualitative explanations which have the 
light from 3C galaxies still dominated
by stars and which ``save'' Lilly and Longair's result. 

The obvious explanation of the third result is if at $\rm z > 2$ we are
seeing radio galaxies that are in the process of formation, because
both the brighter magnitudes and the large dispersion in the K-z
relation 
are what is expected during the epoch of galaxy formation. We have 
discussed
this result further in Eales \& Rawlings (1996).
\bigskip
\centerline{\bf Acknowledgements}
\bigskip

We thank 
James Dunlop for kindly obtaining several images for us, and the
staff at the CFHT and the UKIRT for their support during our various observing
runs, in particular Robin Arsenault, Colin Aspin, Tim Carroll, Dolores
Walther, and Thor Wold. 
We thank the referee, Pat McCarthy, for his helpful comments.
This research has been financially supported
at various stages by the National Science and Engineering Research
Council of Canada and by the Particle Physics and Astronomy Research Council
(PPARC)
of the UK. The United Kingdom Infrared Telescope is operated by the
Royal Observatory Edinburgh on behalf of the PPARC.
\bigskip
\centerline{\bf Refernces}
\bigskip
\parindent = 0pt

\item{} Allington-Smith, J.R. 1982, MNRAS, 199, 611.
\smallskip
\item{} Allington-Smith, J.R., Perryman, M.A.C., Longair, M.S., Gunn,
J.E. \& Westphal, J.A. 1982, MNRAS, 201, 331.
\smallskip

\item{} Baron, E. \& White, S.D.M. 1987, ApJ, 322, 585.
\smallskip

\item{} Barthel, P.D. 1989, ApJ, 336, 606.
\smallskip

\item{} Bruzual, G. \& Charlot, S. 1993, ApJ, 405, 538.
\smallskip

\item{} Bunker, A. 1996, Ph.D. thesis (Oxford), in preparation.
\smallskip

\item{} Burstein, D. \& Heiles, C. 1982, AJ, 87, 1165.

\item{} Chambers, K. C., Miley, 
G. K. \& van Breugel, W. 1987, Nature, 329, 604.
\smallskip

\item{} Charlot, S., \& Bruzual, G.A. 1991, ApJ, 367, 126.
\smallskip

\item{} Cimatti, A., di Serego Alighieri, S., Fosbury,
R.A.E., Salvati, M. \& Taylor, D. 1993, MNRAS, 264, 421.
\smallskip

\item{} Coleman, G.D., Wu, C-C., \& Weedman, D.W. 1980, ApJ suppl.,
43, 393.
\smallskip

\item{} Di Serego Alighieri, S., Cimatti, A. \& Fosbury, R.A.E.
1993, ApJ, 404, 584.

\smallskip

\item{} Dunlop, J. \& Peacock, J. A. 1993, MNRAS, 263, 936.
\smallskip
\item{} Dunlop, J. \& Peacock, J. A., Spinrad, H., Dey, A.,
Jimenez, R., Stern, D. \& Windhorst, R. 1996, Nature, 381, 581.

\smallskip

\item{} Eales, S. A. 1985a, MNRAS, 217, 149. 
\smallskip

\item{} Eales, S.A. 1985b, MNRAS, 217, 167.
\smallskip

\item{} Eales, S. A. 1985c, MNRAS, 217, 179.
\smallskip

\item{} Eales, S. A. 1992, ApJ, 397, 49.
\smallskip

\item{} Eales, S.A. \& Rawlings, S. 1993, ApJ, 411, 67.
\smallskip

\item{} Eales, S.A. \& Rawlings, S. 1996, ApJ, 460, 68.
\smallskip

\item{} Eales, S.A., Rawlings, S., Dickinson, M., Spinrad, H.,
Lacy, M., \& Hill, G. 1993a, ApJ., 409, 578.
\smallskip

\item{} Eales, S.A., Rawlings, S., Puxley, P., Rocca-Volmerage, B.,
\& Kuntz, K. 1993b, Nature, 363, 140.
\smallskip

\item{} Elias, J.H., Frogel, J.A., Matthews, K. \& Neugebauer, G.
1982, AJ, 87, 1029.
\smallskip

\item{} Fanaroff, B.L. \& Riley, J.M. 1974, MNRAS, 167, 31p. 
\smallskip

\item{} Fernini, I., Burns, J.O., Bridle, A.H. \&
Perley, R.A. 1993, AJ, 105, 1690.
\smallskip

\item{} Hill, G.J. \& Lilly, S. 1991, ApJ, 367, 1.
\smallskip

\item{} Jackson, N. \& Rawlings, S. 1996, MNRAS, in press.
\smallskip

\item{} Laing, R.A., Riley, J.M. \& Longair, M.S.
1983, MNRAS, 204, 151.
\smallskip
\item{} Lacy, M. \& Rawlings, S. 1994, MNRAS, 270, 431p.
\smallskip

\item{} Lacy, M., Rawlings, S., Eales, S.A. \& Dunlop, J.S. 1995, MNRAS,
273, 821p.
\smallskip

\item{} Law-Green, J.D.B., Alexander, P.,
Allington-Smith, J.R., van Breugel, W.J.M., Eales, S.A., Leahy,
J.P., Rawlings, S.G. \& Spinrad, H. 1995a, MNRAS,
274, 939.
\smallskip

\item{} Law-Green, J.D.B., Eales, S.A., Leahy,
J.P., Rawlings, S. \& Lacy, M. 1995b, MNRAS,
277, 995.
\smallskip

\item{} Le F\`evre, O., Hammer, F., Nottale, L., Mazure, A., \&
Christian, C. 1988, ApJ, 324, L1.
\smallskip

\item{} Longair, M. S., Best, P.N. \& Rottgering, H.J.A. 1995, 
MNRAS, 275, 47p.
\smallskip

\item{} Lilly, S.J. 1989, ApJ, 340, 77.
 
\item{} Lilly, S. J. \& Longair, M. S. 1984, MNRAS,
211, 833.
\smallskip

\item{} Lilly, S.J., Longair, M.S. \& Allington-Smith, J.R.
1985, MNRAS, 215, 37.
\smallskip

\item{} McCarthy, P.J. 1991, AJ, 102, 518.
\smallskip
\item{} McCarthy, P.J. 1993, ARAA, 31, 639.
\smallskip

\item{}  McCarthy, P. J., Spinrad, H., Djorgovski, S., 
Strauss, M. A., van Breugel, W. \&
Liebert, J. 1987, ApJ, 319, L39.
\smallskip
\item{} McCarthy, P.J., van Breugel, W. J. M. \& Kapahi, V. K. 1991, 
ApJ, 371,
478.
\smallskip
\item{} McCarthy, P.J., van Breugel, W. J. M., Kapahi, V. K.,
\& Subrahmanya, C.R. 1991, AJ, 102, 522. 
\smallskip

\item{} Naundorf, C., Alexander, P., Riley, J.M. \& Eales, S.A.
1992, MNRAS, 258, 647.
\smallskip
\item{} Neeser, M., Eales, S.A., Law-Green, J.D., Leahy, J.P.
\& Rawlings, S. 1995, ApJ, 451, 76.
\smallskip

\item{} Peacock, J.A. \& Wall, J.V. 1981, MNRAS, 194, 331.
\smallskip

\item{} Press, W.H. \& Schechter, P. 1974, ApJ, 187, 425.
\smallskip

\item{} Rawlings, S. et al. 1996, in preparation (Paper II).
\smallskip

\item{} Rawlings, S., Lacy, M. \& Eales, S. 1991, MNRAS, 251, 17p.
\smallskip

\item{} Rawlings, S., Lacy, M., Sivia, D. \& Eales, S., 1995, MNRAS,
274, 428.
\smallskip

\item{} Rawlings, S. \& Saunders, R. D. E. 1991, Nature, 349, 138.
\smallskip

\item{} Rawlings, S., Saunders, R., Eales, S. \& Mackay, C. 1989, 
MNRAS, 240, 701.

\smallskip
\item{} Rawlings, S., Saunders, R., Miller, P., Jones, M.E., \&
Eales, S.A. 1990, MNRAS, 246, 21p.

\smallskip
\item{} Rigler, M.A., Lilly, S.J., Stockton, A., Hammer, F., and
Le F\`evre, O. 1992, ApJ, 385, 61.

\smallskip

\item{} Sandage, A. 1972, ApJ, 173, 485.
\smallskip

\item{} Serjeant, S. et al. 1996, MNRAS, submitted.
\smallskip
\item{} Siegel, S. 1956, {\it Nonparametric statistics for
the behavioural scientists}, (McGraw-Hill).

\item{} Songaila, A., Cowie, L.L., Hu, E.M. \& Gardner, J.P. 1994, ApJ
suppl., 434, 114.
\smallskip

\item{} Spinrad, H., Djorgovski, S., Marr, J., \& Aguilar, L. 1985,
PASP, 97, 932.
\smallskip

\item{} Stockton, A., Kellogg, M. \& Ridgway, S.E. 1995, ApJ, 443, L69.
\smallskip

\item{} Tadhunter, C.N., Fosbury, R.A.E. \& di Serego Alighieri, S.
1988, {\it BL Lac objects: proceedings of the Como conference 1988},
eds Maraschi, L., Maccacaro, T. \& Ulrich, M.-H. (Springer
Verlag), p79.

\smallskip

\item{} Tadhunter, C.N., Scarrott, S.M., Draper, P. \& Rolph, C. 1992,
MNRAS, 256, 53p.
\smallskip

\item{} White, S.D.M., Huchra, J., Latham, D. \& Davis, 
M. 1983, MNRAS, 203, 701.
\smallskip

\item{} Yee, H.K.C. \& Oke, J.B. 1978, ApJ, 226, 753.

\smallskip
\vfill
\eject
\centerline{\bf Figure Captions}
\bigskip
\parindent = 0pt
Figure 1: Radio luminosity at 151 MHz versus redshift for the
3C sample of Laing, Riley \& Longair (1983; open circles) and
for the 6C sample (filled circles). The tight correlation
between radio luminosity and redshift seen for both samples
arises because the samples include only sources which fall
above or between certain flux limits. 
\bigskip

Figure 2: Infrared images of sources in the 6C and B2 samples. North is
always up, east to the left. 
The white crosses on the images mark the positions of radio
components (for astrometric details see the text).
Where the image is an IRCAM or IRCAM3 image the whole field is shown,
a rectangle 52 $\times$ 55 arcsec$^2$ in size in the
case of the IRCAM images and one 89 $\times$ 89 arcsec$^2$ in size
in the case of the IRCAM3 images. Because of the mosaicing technique,
the full sensitivity is only achieved over the central 20 $\times$ 23
arcsec$^2$ for the IRCAM images and over the central 57 $\times$ 57
arcsec$^2$ for the IRCAM3 images. Where the image is a REDEYE image,
only the central 89 $\times$ 89 arcsec$^2$ of the image is shown,
the full sensitivity being achieved over the whole area.
The instrument used to produce each images is as follows:
$\rm 0820+36$ (IRCAM), $\rm 0822+34B$ (IRCAM), $\rm 0822+39$ (REDEYE),
$\rm 0825+34$ (IRCAM), $\rm 0848+34$ (IRCAM), $\rm 0901+35$ (IRCAM),
$\rm 0905+39$ (REDEYE), $\rm 0918+36$ (IRCAM), $\rm 0943+39$ (IRCAM3),
$\rm 1011+36$ (REDEYE), $\rm 1016+36$ (IRCAM), $\rm 1017+37$ (REDEYE),
$\rm 1042+39$ (IRCAM), $\rm 1045+34$ (REDEYE), $\rm 1045+35A$ (IRCAM),
$\rm 1045+35B$ (IRCAM), $\rm 1100+35$ (IRCAM3), $\rm 1123+34$ (REDEYE),
$\rm 1129+37$ (IRCAM3), $\rm 1141+35$ (IRCAM), $\rm 1143+37$ (IRCAM),
$\rm 1204+35$ (IRCAM3), $\rm 1204+37$ (IRCAM), $\rm 1212+38$ (IRCAM),
$\rm 1217+36$ (IRCAM3), $\rm 1230+34$ (IRCAM), $\rm 1256+36$ (REDEYE),
$\rm 1257+36$ (REDEYE), $\rm 1301+35$ (IRCAM).
\bigskip
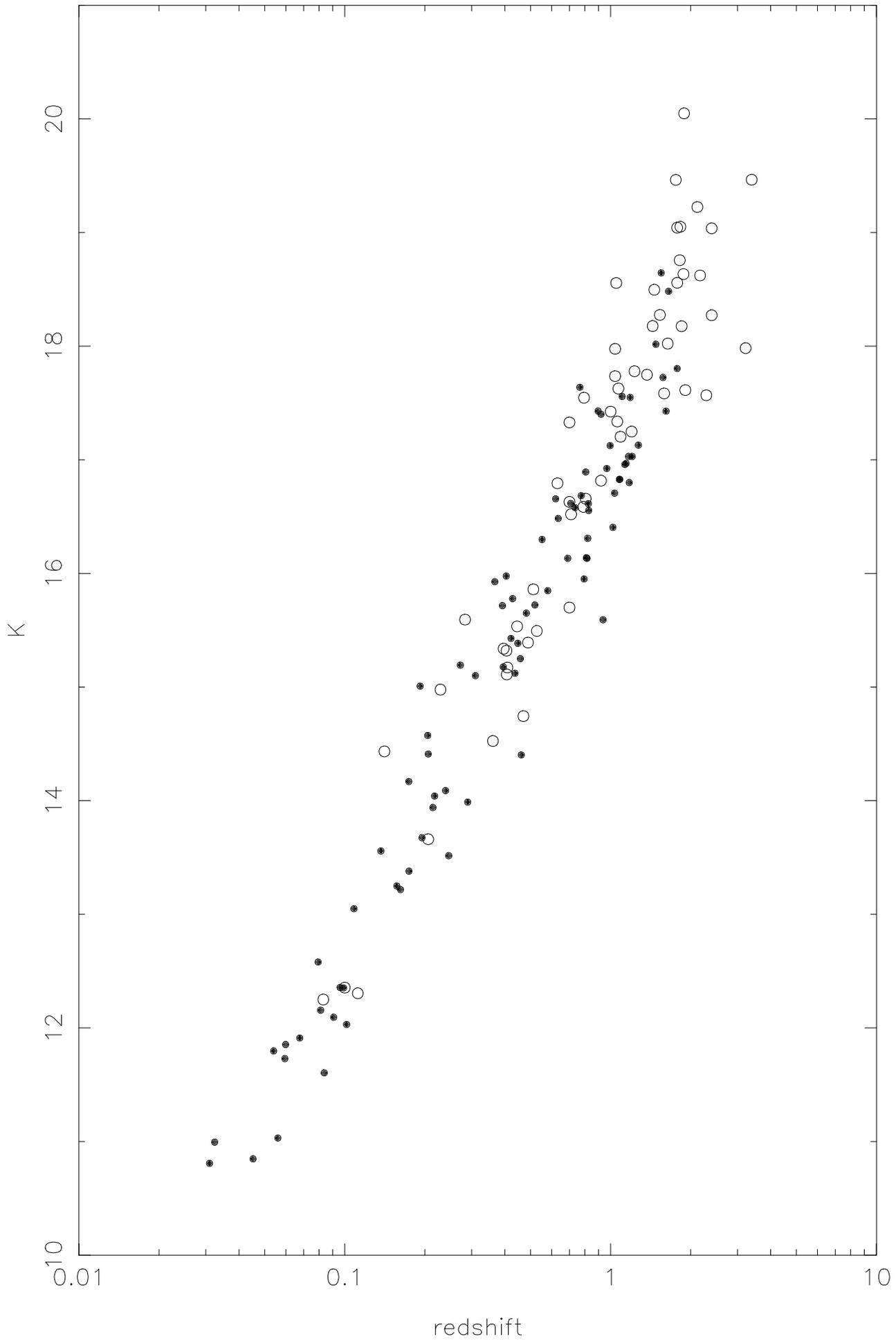
Figure 3: Hubble diagram for the 3C sample (filled circles) and the 6C/B2
sample (open circles). For details of how the diagram has been constructed
see the text.
\bigskip

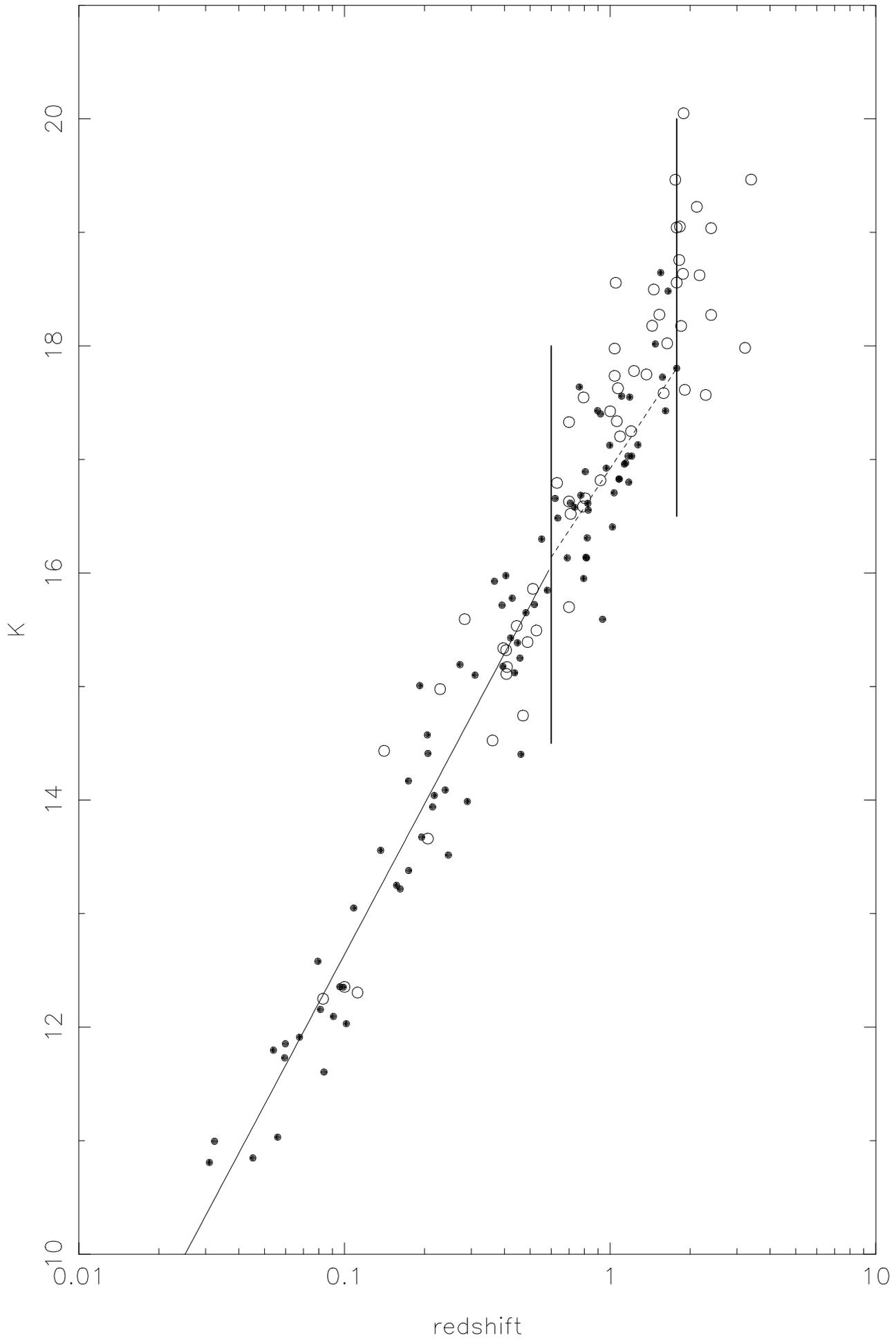
Figure 4: The same Hubble diagram as in Fig. 3 but with the
addition of the results of our simple statistical comparison
of the 3C sample (filled circles) and the 6C/B2 sample (open circles).
The vertical lines show the redshift above which
the alignment effect starts to be seen ($z = 0.6$ - McCarthy 1993)
and the maximum redshift of galaxies in 3C ($z = 1.781$). The dashed line shows
our fit to the 3C points in the redshift range $\rm 0.6 < z < 1.782$,
and the continuous line our fit to the 3C points at $\rm z < 0.6$.

\bigskip

Figure 5: Histograms of the difference between the magnitude of a galaxy
and the best-fitting line in Fig. 4. Figures 5(a) and 5(b) are for the
3C and 6C galaxies in the low-redshift ($\rm z < 0.6$) bin, and Figures
5(c) and 5(d) are for the
3C and 6C galaxies in the high-redshift ($\rm 0.6 < z < 1.8$) bin.

\bigskip

Figure 6: Histogram of the absolute difference between K-band and radio
position angle for B2/6C galaxies with the FR2 radio structure in
the redshift range $\rm 0.8 < z < 1.5$.

\bigskip

Figure 7: Same as Fig. 3 except that the curves show the relations predicted,
first,
if radio galaxies do not evolve and, second, for
various assumptions about their star-formation histories.
The diagram in Fig. 7a is for
$\rm \Omega_0
= 1$ and that in Fig. 7b for $\rm \Omega_0 = 0$.
The continuous lines in both diagrams are the curves expected
if radio galaxies show no evolution (see footnote for details
of how this is calculated).
The other lines show the relations predicted if all the stars
in a radio galaxy form at a continuous rate for a fixed period and
if there is no star formation after this period.
The lowest line is for a model in which the
star formation starts at $\rm z = 10$ and lasts 1 Gyr,
the middle line for one in which the star formation starts
at $\rm z = 100$ and lasts 1 Gyr, and the highest line for
one in which the star formation starts at $\rm z = \infty$ and
lasts 0.1 Gyr. The last model gives the least possible evolution
between now and $\rm z \sim 1$ for a single-burst
model,
since 0.1 Gyr is a typical galactic
free-fall time and is thus the minimum period in which all
the stars can form. Details of how these curves are calculated are given
in the text. The triangle shows the position of the high-redshift
galaxy with a low radio luminosity observed by Dunlop et al. (1996).
\vfill
\supereject
\end


\begin{center}
\begin{tabular}{|l|l|l|l|l|l|l|l|}\hline
(1) & (2) & (3)& (4) & (5) & (6) &(7) & (8)  \\
Name & IRCAM(5) & IRCAM(8) & IRCAM3(5) & IRCAM3(8) & Red(5) & Red(8) & Lilly \\ \hline\hline
0820$+$36 SW &  $\rm 19.00\pm0.26$ & $\rm 18.40\pm0.25$ & ... & ... & ... & ... & ... \\
0820$+$36 NE* & $\rm 18.41\pm0.16$ & $\rm 18.16\pm0.21$ & ... & ... & ... & ... & ... \\  
0822$+$34B NW & $\rm 18.27\pm0.13$ & $\rm 17.57\pm0.12$ & ... & ... & ... & ... & ... \\
0822$+$34B SE* & $\rm 17.71\pm0.09$ & $\rm 17.33\pm0.10$ & ... & ... & ... & ... & ... \\
0822$+$39 & ... & ... & ... & ... & $\rm 17.30\pm0.06$ & $\rm 17.22\pm0.06$ & $\rm 17.39\pm0.11$ \\
 & & & & & & & \\
0825$+$34 NW & $\rm 19.19\pm0.21$ & $\rm 18.51\pm0.19$ & ... & ... & ... & ... & ... \\
0825$+$34 SE* & $\rm 19.00\pm0.18$ & $\rm 18.47\pm0.19$ & ... & ... & ... & ... & ... \\
0848$+$34 & $\rm 15.83\pm0.06$ & $\rm 15.70\pm0.06$ & ... & ... & ... & ... & ... \\
0901$+$35 SW & $\rm 18.48\pm0.17$ & $\rm 18.15\pm0.21$ & ... & ... & ... & ... & ... \\
0901$+$35 NE* & $\rm 18.29\pm0.14$ & $\rm 17.60\pm0.13$ & ... & ... & ... & ... & ... \\
 & & & & & & & \\
0905$+$39 & $\rm 18.48\pm0.17$ & ... & ... & ... & $\rm 19.11\pm0.12$ & ... & ... \\
0918$+$36 & $\rm 16.88\pm0.14$ & $\rm 16.52\pm0.22$ & ... & ... & ... & ... & ... \\
0943$+$39 & $\rm 18.01\pm0.09$ & $\rm 17.94\pm0.13$ & $\rm 18.09\pm0.09$ & $\rm 17.95\pm0.13$ &
$\rm 18.13\pm0.12$ & $\rm 18.13\pm0.13$ & ...\\
1011$+$36 & $\rm 18.05\pm0.14$ & $\rm 17.90\pm0.20$ & ... & ... & $\rm 17.73\pm0.06$ & $\rm 17.51\pm0.07$ &
... \\
1016$+$36 &  ($\rm 20.3^{+0.3}_{-0.2}$)& ... & ... & ... & ... & ... & ... \\
 & & & & & & & \\
1017$+$37 & $\rm 18.88\pm0.19$ & $\rm 18.85\pm0.29$ & ... & ... & $\rm 18.54\pm0.12$ & $\rm 18.29\pm0.13$ &
... \\
1042$+$39 & ($\rm 19.7^{+0.5}_{-0.3}$) & ... & ... & ... & ... & ... & ... \\
1045$+$34 & $\rm 18.91\pm0.17$ & ... & ... & ... & $\rm 18.93\pm0.08$ & ... & ... \\
1045$+$35A & $\rm 16.33\pm0.06$ & $\rm 16.30\pm0.07$ & ... & ... & ... & ... & ... \\
1045$+$35B & $\rm 16.77\pm0.06$ & $\rm 16.63\pm0.07$ & ... & ... & ... & ... & ... \\
 & & & & & & & \\
1100$+$35 & ... & ... & $\rm 18.33\pm0.27$ & ... & ... & ... & ($\rm 17.11\pm0.18$) \\
1123$+$34 & $\rm 17.92\pm0.11$ ... & ... & ... & $\rm 17.81\pm0.11$ & $\rm 17.75\pm0.12$&
... \\
1129$+$37 & ... & ... & $\rm 17.52\pm0.06$ & $\rm 17.31\pm0.07$ & $\rm 17.43\pm0.06$ & $\rm 17.34\pm0.07$ &
($\rm 17.15\pm0.15$)\\
1141$+$35 & $\rm 18.72\pm0.14$ & ... & ... & ... & ... & ... & ... \\
1143$+$37$^{\ddag}$ & ($\rm 19.83^{+0.56}_{-0.37}$) & ... & ... & ... & ... & ... & $\rm 18.97\pm0.35$ \\
\hline
\end{tabular}
\end{center}
\vfill
\eject
\begin{center}
\begin{tabular}{|l|l|l|l|l|l|l|l|}\hline
1204$+$37 & $\rm 19.29\pm0.13$ & ... & ... & ... & ... & ... & ... \\
1204$+$35 & ... & ... & $\rm 17.97\pm0.07$ & $\rm 17.72\pm0.10$ & $\rm 17.88\pm0.06$ & $\rm 17.79\pm0.07$
& ($\rm 17.80\pm0.23$)\\
1212$+$38 NW & $\rm 17.44\pm0.06$ & $\rm 17.15\pm0.07$ & ... & ... & ... & ... & ... \\
1212$+$38 SE* & $\rm 17.74\pm0.07$ & $\rm 17.35\pm0.08$ & ... & ... & ... & ... & ... \\
1217$+$36 & ... & ... & $\rm 17.27\pm0.05$ & 
$\rm 17.02\pm0.06$ & $\rm 17.28\pm0.11$ & $\rm 17.17\pm0.12$ &
($\rm 17.14\pm0.16$) \\   
 & & & & & & & \\
1230$+$34 & $\rm 18.40\pm0.14$ & $\rm 18.25\pm0.21$ & ... & ... & ... & ... & ... \\
1256$+$36 & ... & ... & $\rm 17.75\pm0.06$ & $\rm 17.60\pm0.08$ & $\rm 17.52\pm0.06$ & $\rm 17.10\pm0.06$ &
$\rm 17.61\pm0.15$ \\
1257$+$36 & $\rm 17.51\pm0.08$ & $\rm 17.43\pm0.11$ & ... & ... & $\rm 17.58\pm0.12$ & $\rm 17.36\pm0.12$ &
... \\
1301$+$35 & ($\rm 19.46^{+0.35}_{-0.27}$) & ... & ... & ... & ... & ... & $\rm 18.02\pm0.25$ \\
\hline
\end{tabular}
\end{center}
\parindent = 0pt
The format of this table is as follows. Col. (1) gives the name of the object or the component of the
object. If there are two component listed, there
are two objects in the infrared image close to
the radio position (Fig. 2). The asterisk indicates
the component we have concluded is the site of the
active nucleus. Cols (2)-(7) give the magnitudes measured with IRCAM, IRCAM3 and REDEYE. The name
at the head of each column gives the instrument with which the object was observed and the  
number in brackets is the diameter in arcsec through which the magnitude was measured, unless
there are brackets around the magnitude itself---in which case the diameter of the aperture
used was 4 arcsec. Col. (8) gives the magnitude measured either by Lilly (1989) or by Lilly,
Longair \& Allington-Smith (1985). This was measured through an aperture of diameter
8 arcsec, unless the magnitude is in brackets---in which case the aperture had a diameter
of 12 arcsec. 
\medskip

$\ddag$ The object is significant at the 2.6$\sigma$ level.
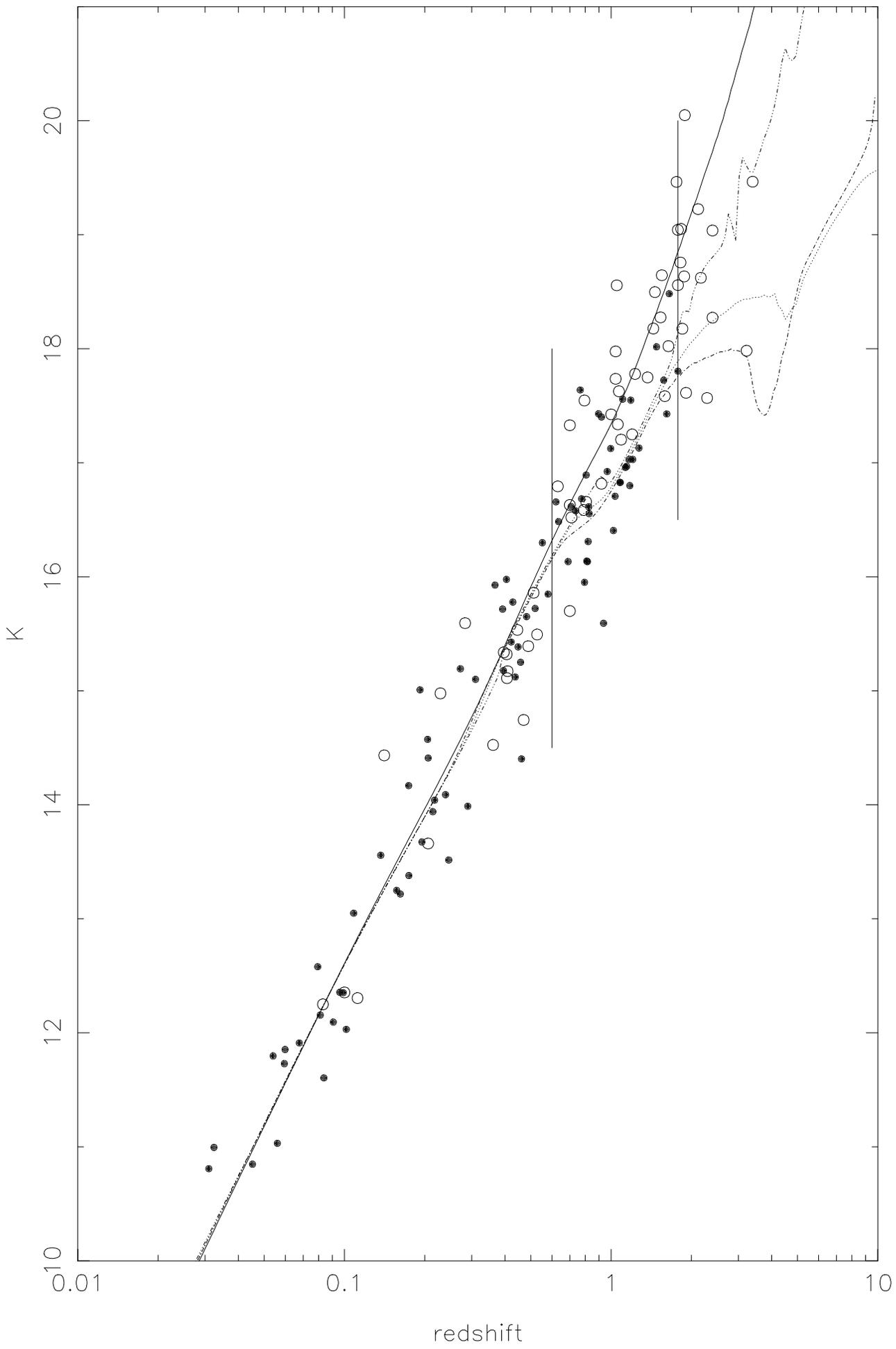
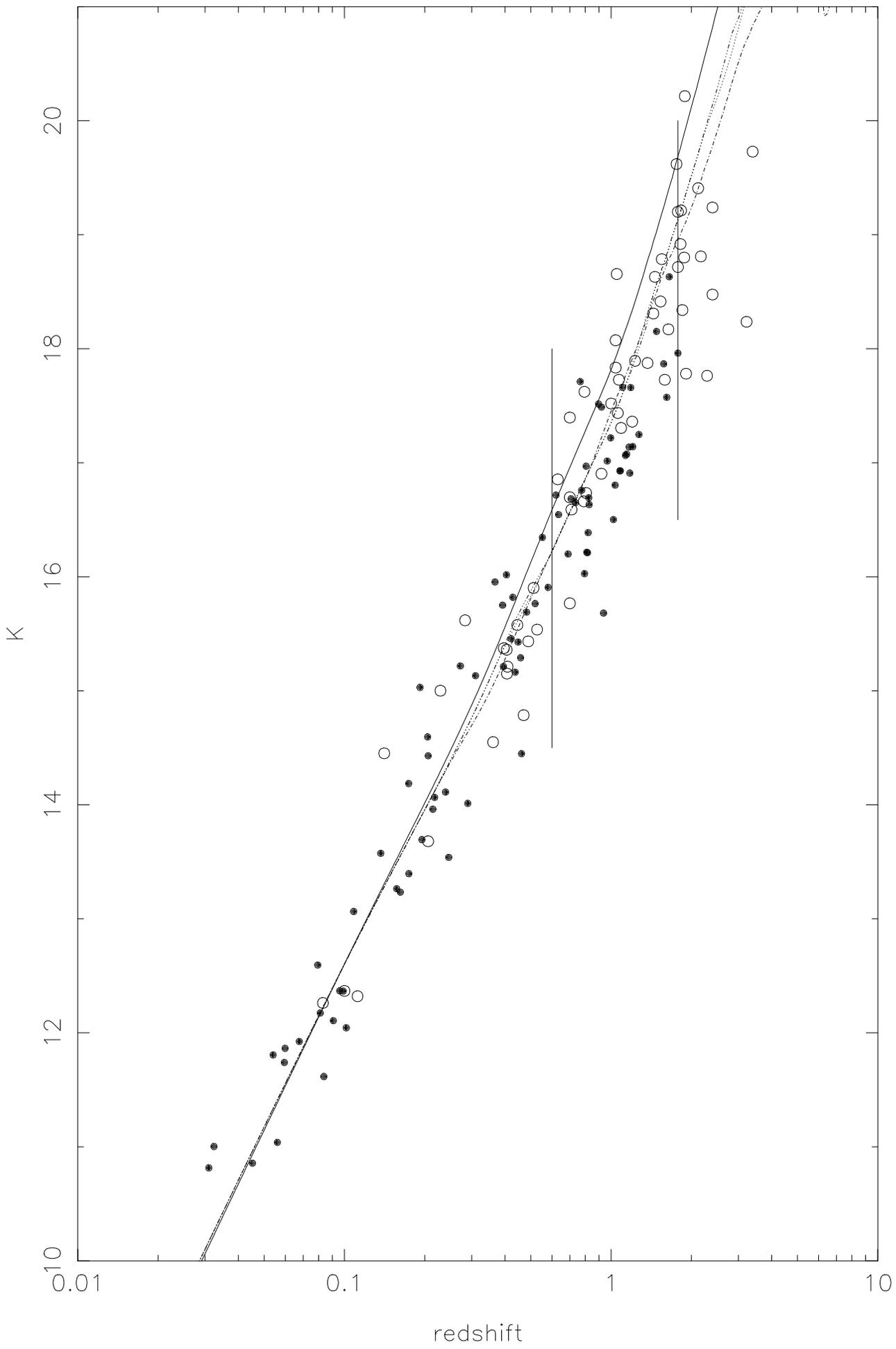
